\documentstyle[prl,aps,epsf]{revtex}
\draft
\begin{document}
\twocolumn[\hsize\textwidth\columnwidth\hsize\csname @twocolumnfalse\endcsname

\title{Local Melting and Drag 
for a Particle Driven Through a Colloidal Crystal} 
\author{C. Reichhardt and C.J. Olson Reichhardt} 
\address{ 
Center for Nonlinear Studies and Theoretical Division, 
Los Alamos National Laboratory, Los Alamos, NM 87545}

\date{\today}
\maketitle
\begin{abstract}
We numerically investigate a colloidal particle driven through
a colloidal crystal as a function of temperature. When the charge of 
the driven particle is larger or comparable to that of the colloids 
comprising the crystal, a local melting can occur, characterized by 
defect generation in the lattice surrounding the driven particle. 
The generation of the defects is accompanied by an 
increase in the drag force on the driven particle, as well as large 
noise fluctuations.  We 
discuss the similarities of these results to 
the peak effect phenomena observed for vortices in superconductors. 
\end{abstract}
\vspace{-0.1in}
\pacs{PACS numbers: 82.70.Dd, 74.25.Qt}
\vspace{-0.3in}

\vskip2pc]
\narrowtext
Colloidal crystals are an ideal system    
for studying a variety of issues 
that arise in two-dimensional (2D) 
systems such as melting \cite{Grier}, defect dynamics  \cite{Ling}, 
and ordering on 2D \cite{Bechinger} and 1D 
\cite{Leiderer} periodic substrates. 
A particular advantage of this system
is that, due to the particle size,
the individual colloid positions and motions can
be directly observed, unlike other systems in which this
information is generally inaccessible. 

Recently there has been growing interest in controlling colloids
individually or in small groups by means of optical techniques 
such as holographic optical trap arrays \cite{Korda}. 
With such methods, individual colloids can be captured
and moved, or collections of colloids can be driven 
through single traps or periodic arrays of traps \cite{Koss,Curtis,Korda2}.
Alternative methods for driving individual colloids include 
moving single magnetic particles through assemblies of non-magnetic
colloids \cite{Weeks}. 
These methods of manipulating and driving individual colloids 
offer a wealth of new ways to explore  
the {\it dynamical}
properties of colloidal crystals and glasses, and  
can also be used to manipulate particles in
other systems such as dusty plasmas \cite{Goree}

A relatively simple 
example of manipulating single colloids to probe collective
properties of a surrounding colloidal crystal  
is to drive a single colloid through a 
crystalline array of other colloids.
One question that arises in this system is
how the size of the driven particle affects both its motion
and the response of the surrounding colloids. 
A very small driven particle is unlikely to 
generate enough stress in the surrounding lattice to 
create defects, and therefore the driven particle motion
and the response of the system will be elastic.
As the particle size or the temperature is increased,
defect generation
may become possible and a transition to plastic flow can occur.
The defects may remain localized near the driven particle 
and strongly affect the frictional drag.
The drag should also depend on the orientation of the driving
direction with 
respect to the colloidal lattice. For example, in a triangular lattice,
easy flow directions should occur along 60$^\circ$ angles.
Work on systems of particles flowing over different orientations
of rigid substrates has shown
that a series of magic angles can arise where the flow locks
into particular orientations \cite{Curtis}. 
 
By studying colloids moving through a periodic substrate
created by other colloids, it may be possible to gain insight into 
related phenomena such as the role played by dislocations
in friction and depinning.
For example, in the case of atomic friction
\cite{Persson} where one lattice is driven over another,
dislocations may nucleate in either of the lattices
and can strongly affect the friction coefficient. 
In the case of driven vortices in superconductors, the onset of 
dislocations or plasticity at 
or near melting can give rise to a phenomena known as the
peak effect, where the vortices slow 
considerably or become pinned 
as a function of external magnetic field or temperature
\cite{Higgins,LingP,Forgan,Zimanyi}.  

In order to investigate the dynamics of a single driven colloidal particle 
moving through a triangular colloidal lattice we 
conduct a series of Langevin simulations
of the type employed in Refs. \cite{Reichhardt,Hastings}. 
We consider a 2D system 
of $N$ colloidal particles interacting with a repulsive Yukawa potential in the
absence of a substrate.  We impose periodic boundary conditions in the $x$ and
$y$ directions. The initial configuration of the colloids is a triangular
lattice.  For a fixed density $n$ there is a well defined 
melting temperature $T_m$, measured either by
diffusion or by the density of 
non-sixfold coordinated particles as determined by a Voronoi tessellation. 
An additional colloid is placed in the system, and a constant 
driving force $f_{d}$ is applied only to that particle.  

 The equation of motion for 
a colloid $i$ is  
\begin{equation}
\frac{d {\bf r}_{i}}{dt} = {\bf f}_{ij} + {\bf f}_{d} + {\bf f}_{T}
\end{equation}
Here ${\bf f}_{ij} = -\sum_{j \neq i}^{N_{c}}\nabla_i V(r_{ij})$ 
is the interaction force from the other colloids.  
The colloid-colloid interaction
is a  Yukawa or screened Coulomb
potential given by
$V(r_{ij}) = (q_{i}q_{j}/|{\bf r}_{i} - {\bf r}_{j}|)
\exp(-\kappa|{\bf r}_{i} - {\bf r}_{j}|)$, 
where  $q_{i(j)}$ is the particle charge, $1/\kappa$ is the 
screening length, and ${\bf r}_{i(j)}$ is the position of
particle $i (j)$.
All the particles have the same charge $q$ except for the driven particle
which has charge $q_{d}$ that may differ from $q$.
The applied driving force ${\bf f}_{d}$ is finite 
if colloid $i$ is the driven colloid and zero otherwise.
The system size is measured in units of the lattice constant 
$a_{0}$ and we use
a screening length of $\kappa = 3/a_{0}$.
The thermal force ${\bf f}_{T}$ is a randomly fluctuating force 
from random kicks. For most of the results in this work we 
fix 
the colloid density (thus fixing the melting temperature 

\begin{figure}
\center{
\epsfxsize=3.5in
\epsfbox{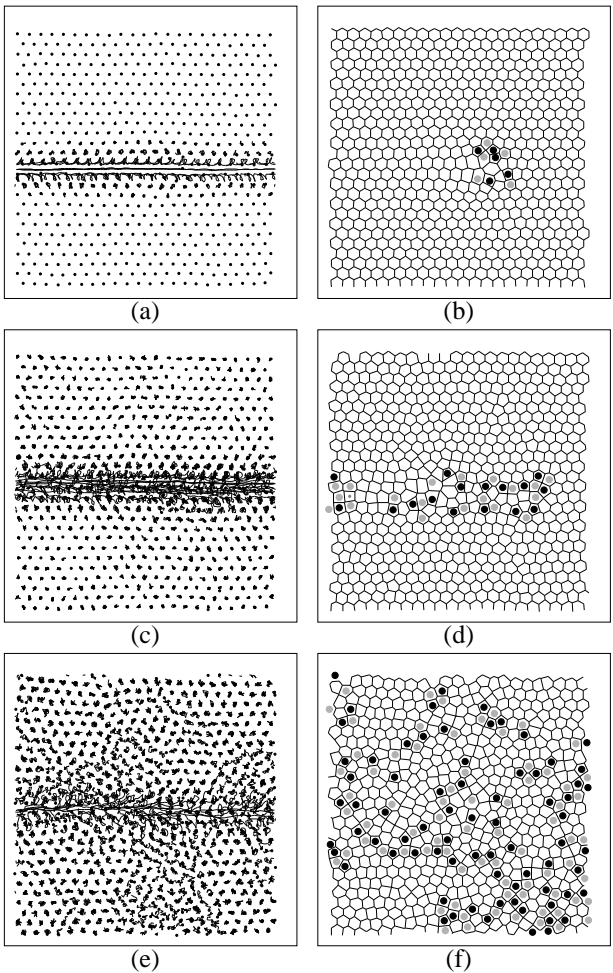}}
\caption{
(a,c,e): Colloid configurations (black dots) 
and trajectory lines at different
temperatures for a driven particle with charge $q_{d}/q = 3$.  
(b,d,f): Corresponding Voronoi construction
colored according
to the number of neighbors: five (black), six (white), and seven (gray).
(a,b) $T/T_{m} = 0.1$; (c,d)
$T/T_{m} = 0.57$; (e,f) $T/T_{m} = 1.07$. 
}
\end{figure}

\noindent
$T_{m}$) and 
vary $T$ and $q_{d}$. For results with constant $f_{d}$
we use the same drive for all $q_{d}$. In the absence of any other
colloids the driven particle moves at a velocity $V_{0}$.    

We first consider the case where a particle with $q_{d}/q=3$ is driven along
the zero degree angle with respect to the lattice. 
Three distinct phases occur for increasing $T/T_{m}$. In Fig.~1(a,c,e) we 
show the colloids 
and particle trajectories for different $T/T_{m}$. 
In Fig.~1(b,d,f) we show the corresponding Voronoi 
construction or Wigner-Seitz cells, in which a single colloid is located
at the center of each polygon. The cells are colored according to the number
of nearest neighbors: 
white for six neighbors, black for five
neighbors, and gray for seven neighbors.
In Fig.~1(a,b) at $T/T_{m} = 0.1$, the flow is elastic and
the particle moves along a 1D path while causing small 
distortions in the surrounding
lattice. Fig.~1(b) shows that although a group of defects
surrounds the driven particle, which
is an extra or interstitial particle in the 

\begin{figure}
\center{
\epsfxsize=3.5in
\epsfbox{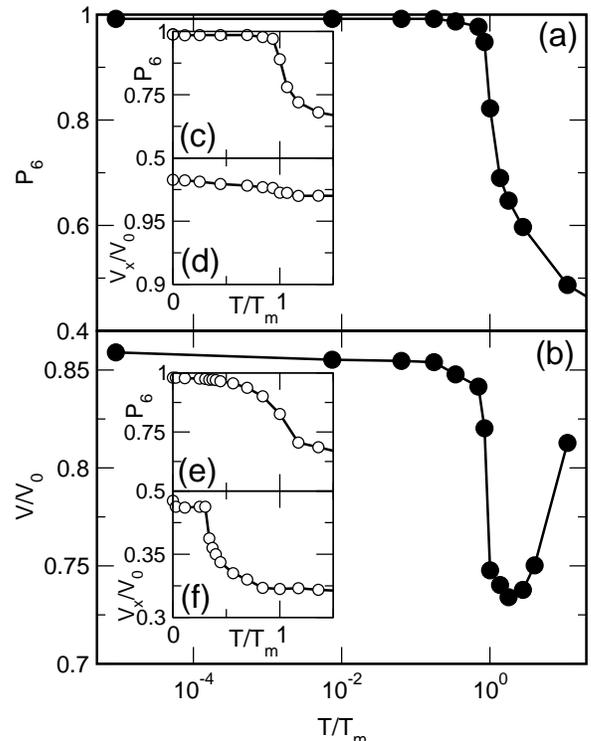}}
\caption{(a,c,e) Fraction of colloids with six neighbors $P_{6}$ vs $T/T_{m}$;
(b,d,f) corresponding velocity $V_x$ in the direction of drive.
(a,b) $q_{d}/q=1$; 
(c,d) $q_{d}/q = 0.25$; 
(e,f) $q_{d}/q = 6$.
}
\end{figure}

\noindent
triangular lattice,
there are no other dislocations induced elsewhere
in the lattice.
At $T/T_{m} = 0.3$, we find a transition to plastic flow 
where the particle no longer moves strictly in a straight line, and 
defects form in the surrounding lattice. Additionally, some 
colloids from the surrounding lattice are pushed in front of the driven
particle over distances larger than a lattice constant $a_0$. Fig.~1(c,d) 
illustrates
this behavior for $T/T_{m} = 0.57$.  A localized molten 
region surrounds the moving particle, and ejects 5-7 defects 
into the lattice over time. 
Fig.~1(e,f) shows the system above melting at $T/T_{m} = 1.07$, 
where defects are generated in the sample even in the absence of the
driven particle.  In this globally molten phase, the driven particle
is still able to push other particles in front of it.    

We next show the effect of the defect generation on the
colloidal velocity in the different
phases. In Fig.~2(a) we plot the percentage of sixfold coordinated particles,
$P_6$, for a system with $q_{d}/q = 1$, fixed $f_{d}$,
and increasing temperature.
We find a transition from  elastic to plastic flow
at a temperature close to $T/T_{m}$. 
In Fig.~2(b), we see that the corresponding colloid velocity 
$V_{x}<V_{0}$ even for $T = 0$, indicating that
some of the energy of the driven particle is absorbed by vibrations in the
surrounding lattice. As $T$ increases, $V_x$ decreases slightly,
culminating in a sharp drop at $T/T_{m} = 1$ which coincides with
the drop in $P_{6}$.  We also find a noticeable decrease in 
$V_{x}$ at temperatures slightly below $T_m$ which is 
correlated with a decrease in $P_{6}$, 
indicating that the defects in the colloidal
lattice slow the driven particle. As $T$ is further increased,
$V_{x}$ eventually increases to $V_0$ for $T/T_m>10.$

We next consider the effect of varying $q_{d}$.
In Fig.~2(c) we plot $P_{6}$ for a system with $q_{d}/q = 0.25$,
and in Fig.~2(d) the corresponding $V_{x}$. Here $V_{x}$ is only
slightly lower than $V_{0}$. A small decrease in $V_{x}$ 
occurs close to melting; however, 
unlike the case of $q_{d}/q = 1$, there is 
little change in $V_{x}$ across $T_{m}$. 
In Fig.~2(e) we show 
$P_{6}$ for $q_{d}/q = 6$, 
and in Fig.~2(f) the corresponding
$V_{x}$. 
$P_{6}$ is more rounded than for $q_{d}/q = 0.25$.
This is due to the dislocations created by the driven particle,
which become more numerous as the melting transition is approached. 
$V_{x}$ drops dramatically well below the
melting temperature, at a much lower $T$ than
observed for smaller $q_{d}$. The decrease in $V_x$ coincides with the
creation of dislocations in the lattice near the driven 
particle. Far from the driven particle, the colloidal lattice remains
ordered. 

We observe 
that at the onset of the local dislocation or defect creation, the driven 
particle often drags an additional colloid in front of it. Once the
dragged colloid falls away to one side, another colloid is captured
and dragged.
The presence of an extra dragged particle produces extra 
resistance to the motion of the driven particle.
In Fig.~2(b), at high temperatures $T \gg T_m$, the drag effect is reduced
and the particle velocity increases 
when the thermal fluctuations become so strong that the driven 
particle can no longer capture other particles. 
Small driven particles do not have a strong enough colloid-colloid interaction 
to push any additional particles, so no additional drag effect occurs. 	
The small driven particle of Fig.~2(c) creates
little distortion in the surrounding colloidal lattice, 
so the formation of dislocations occurs only due to 
$T$. In  Fig.~2(e) the driven particle is large enough to create a significant
amount of distortion in the surrounding lattice, and thus
produce dislocations at low $T$ well below melting. 
Thus, the {\it combination} of lattice distortions from
the driven particle and the temperature produces the
onset of dislocations. Dislocations appear at a lower temperature
in the presence of a large driven particle than they do for a small or
for no particle, when no lattice distortions are introduced.
We therefore expect that as $q_d$ increases, 
the local melting transition will shift to lower $T$.  
We have also considered the effect of varying $f_{d}$, 
and observe the same results whenever
the particle velocity is slower than
the propagation of disturbances in the surrounding media.

In Fig.~3 we plot the curve separating the crystal phase C,
the local melting phase LM, and the global melting phase M. 
The line separating C and LM is obtained from the decrease of $V_{x}$. 
Within LM, the moving particle creates dislocations but only locally
as seen in Fig.~1(d). 

We note that these results have many similarities with 
a phenomena called the 
peak effect for vortices in superconductors 
\cite{Higgins,LingP,Forgan,Zimanyi}.
An immediate difference between the two systems is that
the vortices interact with a background
of quenched disorder which is not present in our 

\begin{figure}
\center{
\epsfxsize=3.5in
\epsfbox{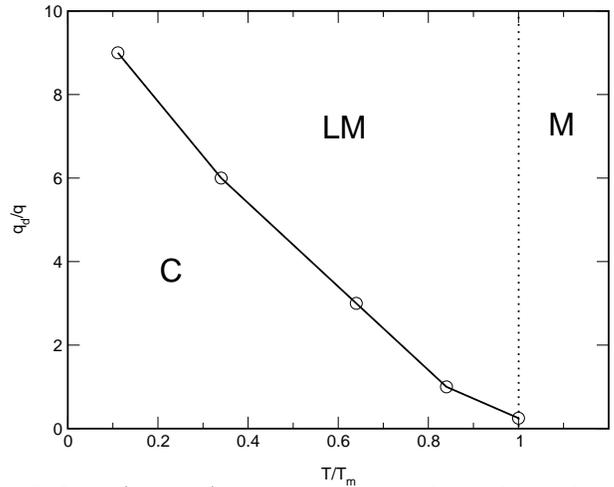}}
\caption{$q_{d}/q$ vs $T/T_{m}$ for a series of simulations with a constant 
drive. C is the crystal phase where there are no defects generated in the
surrounding lattice as in Fig.~1(a,b).  In the local melting phase LM,
defects are generated near the driving particle as in 
Fig.~1(c,d). M is the global melting phase as in Fig.~1(e,f).
}
\end{figure}

\noindent
system. 
The peak effect is generally believed to occur due to a transition
from an elastic, dislocation free vortex lattice to 
a disordered glassy state containing many dislocations. The 
relatively stiff elastic lattice cannot adjust to the disorder and 
is thus weakly pinned, whereas the soft disordered state
can adjust well to the disorder and is strongly pinned. 
The peak effect is observed by monitoring the vortex velocity in
the form of a voltage as a function of temperature.
At the peak effect transition 
the voltage drops abruptly, indicating that the 
vortices are slowing or being pinned.
The peak effect can also be observed by constructing IV curves.
There is still controversy over whether the true nature of the peak effect
is a thermal melting or a disorder induced transition \cite{LingP,Forgan}. 
Distinguishing between these two scenarios is made difficult by the
fact that the entire vortex lattice becomes disordered at the peak effect
transition.

In our system there is a transition from an elastic (defect free)
flow to a plastic (defected) flow which coincides with a drop
in the velocity of the moving particle. 
The velocity drop occurs even when the defects do not occur 
in the whole sample but only surround the driven particle. 
Additionally we find strong $1/f$ type velocity 
noise fluctuations in the
disordered flow. 
In our system, the velocity transition is not caused by thermal
fluctuations, but is instead due to the distortions of the 
surrounding lattice by the driven particle.
In the case of the vortices, lattice distortions
could arise from the relative motion of vortices 
trapped at pinning sites. If the distortions in the surrounding
lattice are large enough,
dislocations nucleate and the overall vortex velocity drops. 

To further explore similarities between the peak effect and our system, 
we 
slowly increase $f_{d}$ and measure the velocity at fixed $T$,
in analogy with 
IV measurements. A 

\begin{figure}
\center{
\epsfxsize=3.5in
\epsfbox{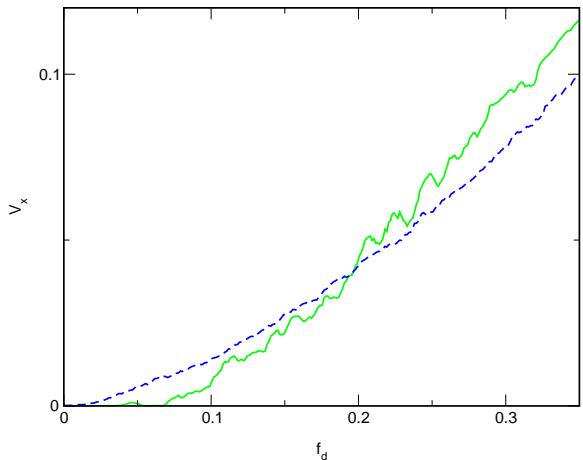}}
\caption{Velocity vs force curve for $q/q_{d} = 3.0$. Solid line: 
$T/T_{m} = 0.5$ in the elastic regime. Dotted line: $T/T_{m} = 1.1$ in
the global melted region.  
}
\end{figure}

\noindent
particular characteristic of the peak effect is 
crossing $IV$ curves \cite{Zimanyi}. 
Due to thermal activation, the more disordered 
system has a lower threshold for motion than the ordered system.
In contrast, at higher drives when thermal creep is negligible, 
the overall velocity is lower in the disordered system than in the
ordered system.  This implies that the velocity-force curves for
the two cases must cross at intermediate drives.

In Fig.~4 we plot the velocity vs force curves for 
$q_{d}/q = 3$ at fixed $T/T_{m} = 0.5$ in the ordered regime (solid line) and
at $T/T_{m} = 1.1$ just into the liquid regime (dotted line). 
For the disordered (melted) system 
there is no true depinning threshold
and we observe a finite velocity almost to zero applied drive.
The velocity in the ordered regime for $f_{d} < 0.2$ 
is {\it less} than that of the disordered regime
and is almost zero, with a clear depinning threshold. 
By contrast, at higher drives 
the velocity in the ordered regime is {\it higher} than the 
melted regime. Thus we observe a crossing of the velocity force
curves  similar the crossing of the IV curves \cite{Zimanyi} for
the peak effect.   
We have also examined the velocity in the local melting case and
again find a finite depinning 
threshold; however, here the velocity is always lower than the
ordered case, even at high drives. 
Our results lend support to the idea that the peak effect phenomena
in superconductors is {\it not} a thermal melting phenomena, but 
is instead a {\it disorder induced transition}.

We have also explored the effect of driving the particle at
various angles with respect to the undriven lattice. For certain angles such 
as $60^{\circ}$ the results are unchanged. For incommensurate angles, the 
plastic flow transition shifts to a lower $T/T_{m}$ since 
the moving particle generates larger distortions in the surrounding
lattice and defect creation is easier.

In conclusion, we have studied 
a single colloid driven through an ordered colloidal crystal. We find
that as a function of temperature and charge of the driven colloid, 
there is a transition from elastic flow, where no defects are generated
in the surrounding lattice, to plastic flow, where defects proliferate. 
This transition coincides with an increased drag on the driven particle, 
due to the driven particle trapping and pushing surrounding colloids
in front of it.
As the charge of the 
driven particle increases, the 
elastic-plastic transition shifts to lower temperatures.  
This system has several similarities to the peak effect phenomena found
in vortex matter in superconductors:  The onset of 
defect formation slows the particle, and there is 
a crossing in the velocity force curves. Our results suggest that
the peak effect is due to a disorder induced transition rather
than a thermal melting transition. 
These results should be accessible 
experimentally by driving single colloids with
optical tweezers or dragging a magnetic bead though an ordered 
colloidal crystal. Another variation on this would be to
drive a colloidal crystal 
past obstacles of varied size. 
Other systems in which these effects could be observed
include driving particles through dusty plasma crystals or
ordered foams.

We thank E.~Weeks and M.B.~Hastings 
for useful discussions. 
This work was supported by the U.S. DoE under Contract No.
W-7405-ENG-36.

\end{document}